\documentclass[aps,pra,twocolumn,superscriptaddress,nofootinbib,longbibliography]{revtex4-2}

\usepackage[utf8]{inputenc}
\usepackage{amsmath,amssymb,bm}
\usepackage{braket}
\usepackage{graphicx}
\usepackage{booktabs}
\usepackage{enumitem}
\usepackage{xcolor}
\usepackage[normalem]{ulem}
\usepackage[colorlinks=true,linkcolor=blue!60!black,citecolor=blue!60!black,urlcolor=blue!60!black,breaklinks=true]{hyperref}

\newcommand{\Tr}{\mathrm{Tr}}
\newcommand{\kp}{\ket{+1}}
\newcommand{\km}{\ket{-1}}
\newcommand{\tp}{\otimes}

\graphicspath{{Figures/}}
\begin{document}

\title{Entanglement swapping through a spin--orbit hybrid qubit interface}

\author{Javier~Fern\'andez}
\thanks{These authors contributed equally to this work.}
\affiliation{Departament de F\'isica, Universitat Aut\`onoma de Barcelona, E-08193 Bellaterra, Spain}
\author{Philipp~Oleynik}
\thanks{These authors contributed equally to this work.}
\affiliation{IPaQS, SUPA, Heriot-Watt University, Edinburgh EH14 4AS, United Kingdom}
\author{Kiki~Dekkers}
\affiliation{IPaQS, SUPA, Heriot-Watt University, Edinburgh EH14 4AS, United Kingdom}
\author{Maddox~Aherin}
\affiliation{IPaQS, SUPA, Heriot-Watt University, Edinburgh EH14 4AS, United Kingdom}
\author{Lorenzo~Marrucci}
\affiliation{Dipartimento di Fisica, Universit\`a di Napoli Federico II, Complesso Universitario di Monte Sant'Angelo, Napoli, Italy}
\author{Bruno~Piccirillo}
\affiliation{Dipartimento di Fisica, Universit\`a di Napoli Federico II, Complesso Universitario di Monte Sant'Angelo, Napoli, Italy}
\author{Filippo~Cardano}
\affiliation{Dipartimento di Fisica, Universit\`a di Napoli Federico II, Complesso Universitario di Monte Sant'Angelo, Napoli, Italy}
\author{Gerard~J.~Machado}
\affiliation{Clarendon Laboratory, University of Oxford, Parks Road, Oxford OX1 3PU, United Kingdom}
\author{Adam~Vall\'es}
\email{adam.valles@uab.cat}
\affiliation{Departament de F\'isica, Universitat Aut\`onoma de Barcelona, E-08193 Bellaterra, Spain}
\author{Andrew~Forbes}
\affiliation{School of Physics, University of the Witwatersrand, Johannesburg 2050, South Africa}
\author{Jonathan~Leach}
\affiliation{IPaQS, SUPA, Heriot-Watt University, Edinburgh EH14 4AS, United Kingdom}

\date{\today}

\begin{abstract}
\noindent\looseness=-1 The fragility of orbital angular momentum (OAM) complicates distributing spatial-mode entanglement over long distances. In both optical fibre and turbulent free space, polarization is far more resilient, but only offers access to two dimensions. We realize an entanglement-swapping node designed to combine the high dimensions of OAM with the resilience of polarization: two OAM-entangled photon pairs are generated by spontaneous parametric down-conversion in a single crystal, and the intermediary photons are mapped onto polarization qubits by $q$-plates and single-mode fibre coupling before interfering at a beam splitter. We certify the swapped entanglement with a witness built from correlations in three mutually
unbiased bases, which exceeds the separable bound by more than nine times its statistical
uncertainty, and the least favourable separable state reaches the measured value with probability
$p\approx3\times10^{-15}$. As the joint measurement acts only on polarization, the node is by design compatible with single-mode fibre and free-space links, and the topological charge of the $q$-plates sets the swapped OAM subspace, offering a route towards multi-dimensional repeater nodes.
\end{abstract}

\maketitle

\section{Introduction}

A global network distributing entanglement between quantum processors, the so-called quantum internet, will require entanglement to be shared reliably over continental distances to enable unconditionally secure communication, networked quantum computation, and distributed sensing beyond classical limits \cite{kimble2008quantum,wehner2018quantum,azuma2023quantum}. Direct photon transmission cannot sustain such distances: transmission in optical fibre decays exponentially with distance, while in free space the beam spreads by diffraction and is distorted by atmospheric turbulence. Quantum repeaters, which divide a long channel into shorter elementary links, were proposed to overcome this scaling limit and remain the standard architecture for long-distance entanglement distribution \cite{briegel1998quantum,azuma2015all, pirandola2017quantum}. Progress towards this goal has followed two routes: satellite links have distributed entanglement over 1200~km by routing most of the path through space, crossing only a thin atmospheric layer near each ground station \cite{yin2017satellite}, while ground-based deployments have distributed polarization-entangled photon pairs continuously over 248~km of deployed telecommunication fibre without trusted nodes \cite{neumann2022continuous}.

\looseness=-1 Encoding in high dimensions promises a larger alphabet per photon and, in principle, greater robustness to noise \cite{cozzolino2019high,erhard2020advances}, with the orbital angular momentum (OAM) of photons providing an in-principle unbounded spatial-mode alphabet \cite{mair2001entanglement,erhard2018twisted,forbes2019quantum,nape2023quantum}. However, transporting OAM modes over long distances is challenging. OAM is poorly preserved by conventional fibre, requiring custom few-mode or air-core designs that have so far limited OAM quantum communication to the kilometre scale \cite{cao2020distribution,cozzolino2019orbital,cozzolino2019air}. Free-space links, in turn, inherit atmospheric turbulence, which scrambles spatial modes. OAM and high-dimensional entanglement have so far been distributed over a few kilometres of urban air \cite{krenn2015twisted,steinlechner2017distribution}. Polarization is far less affected: it is essentially insensitive to turbulence and is preserved in standard single-mode fibre (SMF) up to a slowly drifting unitary, but it is confined to two dimensions. Hybrid spin--orbit encoding \cite{karimi2010spin,nagali2010generation} offers a way around this for point-to-point transport. By entangling the spin of one photon with the orbital degree of freedom (DOF) of its partner, multi-dimensional entanglement has been carried down conventional SMF, with the fragile spatial mode never entering the channel \cite{liu2020multidimensional}.

Entanglement swapping, the elementary operation of a quantum repeater \cite{zukowski1993event,pan1998experimental}, has been demonstrated for spatial modes, including many OAM \cite{zhang2017simultaneous} and pixel-based \cite{bornman2019ghost} subspaces, with the joint measurement in each case performed on the spatial DOF itself. Hybrid entanglement swapping, in which spin--orbit hybrid states allow the Bell-state measurement (BSM) to act on polarization while the swapped entanglement is established in OAM, was proposed theoretically more than a decade ago as a route to OAM Bell and Greenberger--Horne--Zeilinger states \cite{chen2011hybrid}. Polarization qubits have since been mapped onto OAM by remote state preparation and teleportation-based state transfer \cite{erhard2015real,ru2021quantum}, and entanglement swapping between different optical encodings has been realized for discrete- and continuous-variable states \cite{takeda2015entanglement,guccione2020connecting}; to our knowledge, however, spin--orbit hybrid entanglement swapping has not yet been realized experimentally. Repeater nodes that are able to work with larger alphabets, while remaining compatible with the optical fibre already deployed, are a central requirement for extending entanglement distribution beyond currently demonstrated ranges.

\begin{figure*}[t]
\centering
\includegraphics[width=\textwidth]{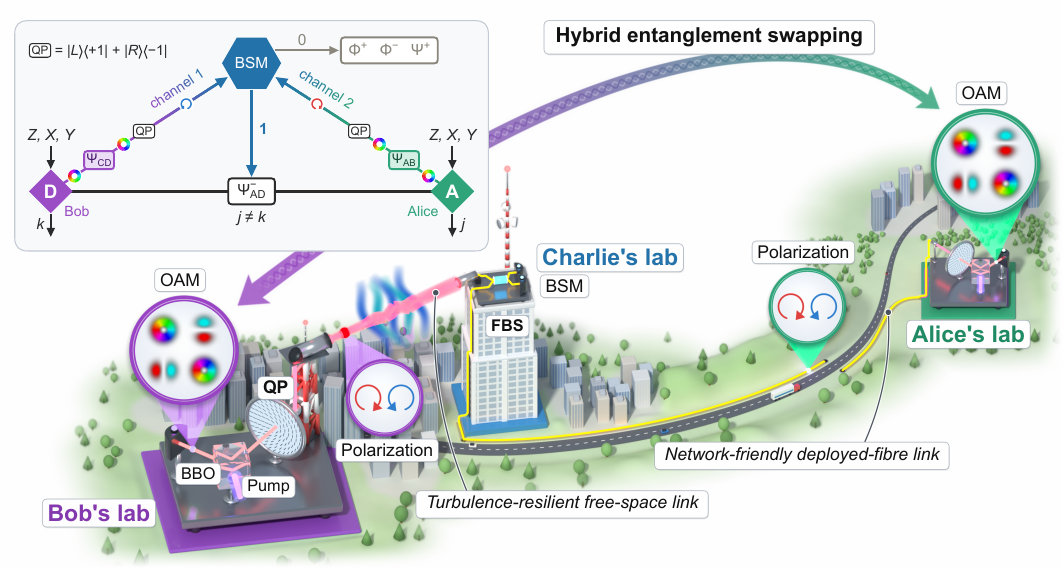}
\caption{\textbf{Future spin--orbit quantum relay with our measurement protocol.} Alice and Bob, in two cities, each keep one photon of an OAM-entangled pair ($A$, $D$) and convert the other ($B$, $C$) to polarization with a $q$-plate (QP), so that OAM stays in the labs and only polarization travels. $C$ reaches Charlie over a free-space link (channel~1) and $B$ over a deployed-SMF link (channel~2); polarization suits both, as it withstands turbulence better than OAM and, unlike OAM, is carried by standard SMF. Charlie's BSM on polarization (FBS and two detectors) projects $A$ and $D$, which never interacted, onto the OAM singlet $\ket{\Psi^-}$, i.e., hybrid entanglement swapping. Top left: our protocol. The four-photon state is $\tfrac12\sum_\beta\ket{\beta}_{AD}\ket{\beta}_{BC}$ over the four Bell states $\beta$; only $\ket{\Psi^-}_{BC}$ (outcome~1) is kept, leaving $A$ and $D$ in $\ket{\Psi^-}_{AD}$, with $j\neq k$ in any common basis.}
\label{fig:concept}
\end{figure*}

Here we demonstrate a proof of principle of such a node in its simplest, memoryless form, a quantum relay: a four-photon experiment in which the most delicate step of the protocol (the two-photon interference at the heart of the swap) is carried out on the robust polarization DOF, which survives both atmospheric turbulence and deployed SMF.
\looseness=-1 In the scenario of Fig.~\ref{fig:concept}, Alice and Bob keep their OAM photons and send only polarization-encoded partners to Charlie, whose joint measurement leaves the photons they kept entangled in OAM. The intermediary photons are transcribed onto polarization qubits by a spin-to-orbit converter, the so-called $q$-plate \cite{marrucci2006optical,nagali2009quantum,marrucci2011spin}, followed by SMF coupling. Their interference at a fibre beam splitter (FBS) heralds, on post-selection of fourfold coincidences, an OAM-entangled singlet between two remote photons that never interacted. We first verify the indistinguishability of the intermediary photons through a remote Hong--Ou--Mandel (HOM) measurement \cite{hong1987measurement}, and then certify the swapped entanglement with a witness built from correlations in mutually unbiased bases (MUBs) \cite{spengler2012MUBs,erker2017quantifying,bavaresco2018measurements}, which requires only two measurement bases per photon (we add a third to increase the statistical significance) and is directly extensible to higher dimensions; a two-qubit full state tomography (FST) of the swapped state is reported for comparison. Since only the intermediary arms pass through the spin--orbit interface, the remote photons are not restricted by it: by OAM conservation, a swapping event projects them onto the two-dimensional subspace $\ell=\pm2q$ selected by the topological charge $q$ of the $q$-plates, which the user can therefore choose.

\section{Concept and experimental implementation}
\label{sec:setup}
Two OAM-entangled photon pairs, $A$--$B$ and $C$--$D$, are generated by spontaneous parametric down-conversion (SPDC) in a single crystal [Fig.~\ref{fig:setup-data}(a); a detailed layout of the setup is given in the Supplemental Material (SM), Fig.~\ref{fig:setup-detailed}]. The remote photons $A$ and $D$ are projected onto the two-dimensional OAM subspace spanned by $\ell=+1$ and $\ell=-1$, which we label $\kp$ and $\km$. The polarization basis $\{\ket{H},\ket{V}\}$ and the circular basis $\{\ket{R},\ket{L}\}$ are used for the polarization-encoded intermediary photons. In the OAM subspace we use the three MUBs
\begin{align}
    Z&=\{\ket{z_0},\ket{z_1}\}=\{\kp,\km\},\nonumber\\
    X&=\{\ket{x_0},\ket{x_1}\}=\{(\kp\pm\km)/\sqrt2\},\label{eq:bases}\\
    Y&=\{\ket{y_0},\ket{y_1}\}=\{(\kp\pm i\km)/\sqrt2\}.\nonumber
\end{align}
Spatial light modulators (SLMs) followed by SMFs placed in the $A$ and $D$ arms perform projective measurements onto these states, selecting the subspace in which entanglement is certified.

The intermediary photons $B$ and $C$ each pass a $q$-plate \cite{marrucci2006optical}, which couples the spin (circular polarization) of the photon to its OAM according to the conversion rules
\begin{equation}
    \ket{R,\ell}\mapsto\ket{L,\ell-2q},\qquad \ket{L,\ell}\mapsto\ket{R,\ell+2q},
    \label{eq:qplate-ladder}
\end{equation}
where $q$ is the topological charge of the plate; here $q=1/2$. The photons are then coupled into SMF, which transmits only the Gaussian ($\ell=0$) component, so that the conversion leaves only a \emph{polarization} label that encodes the selected OAM value,
\begin{equation}
    \kp\mapsto\ket{L},\qquad \km\mapsto e^{i\chi}\ket{R},
    \label{eq:map}
\end{equation}
with the sign and the phase $\chi$ fixed by calibration. In this way, the OAM qubit carried by $B$ and $C$ before the $q$-plate is mapped one-to-one onto a polarization qubit after the SMF. These two fibres are the inputs of an FBS, and the polarization is controlled so that the two inputs are matched. There, a valid swapping event is recorded as a fourfold coincidence between the two outputs and the two remote detectors, conditioned on the OAM projections displayed on the SLM$_{A}$ and SLM$_{D}$ masks. As $A$ and $D$ are detected destructively, the swapped state is characterized on these post-selected fourfold events.

A coincidence event with one photon in each output of the FBS acts on the intermediary polarizations as a projector onto the antisymmetric Bell state $\ket{\Psi^-}_{BC}$, one of the four Bell states, so that the BSM succeeds with probability $1/4$ in the absence of losses, and the remote photons collapse onto the OAM singlet
\begin{equation}
    \ket{\Psi^-}_{AD}=\frac{\kp_A\km_D-\km_A\kp_D}{\sqrt{2}},
    \label{eq:AD-target}
\end{equation}
up to a phase that is fixed once in calibration. The derivation, starting from the source states and following the four-photon state through the spin--orbit map and the FBS projection, is given in Sec.~\ref{sec:methods}.

\begin{figure*}[t]
\centering
\includegraphics[width=\textwidth]{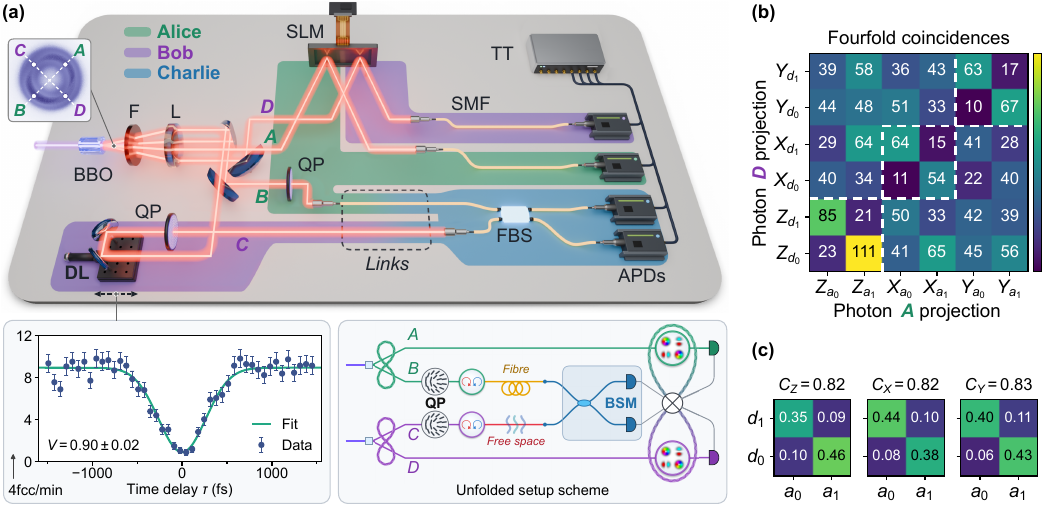}
\caption{\looseness=-1 \textbf{Experimental implementation and data.} (a) Simplified view of the setup (not to scale; see Sec.~\ref{sec:setup-methods} and the SM, Fig.~\ref{fig:setup-detailed}), with the paths of Alice, Bob and Charlie on coloured regions. Insets: camera image of the SPDC ring with the four collected points, remote HOM dip with the $q$-plates off (4fcc/min: fourfold coincidences per minute; 10~h of acquisition in total) and unfolded scheme of the setup. BBO: beta-barium borate crystal; F: filter; L: lens; QP: $q$-plate; DL: delay line; TT: time tagger; APDs: avalanche photodiodes. (b) Fourfold coincidences for the 36 projections of photons $A$ and $D$ onto the MUB states of Eq.~\eqref{eq:bases}; the dashed lines delimit the $Z$, $X$ and $Y$ blocks. (c) These blocks, each normalized within its basis, with the anti-diagonal weights $C_Z$, $C_X$ and $C_Y$.}
\label{fig:setup-data}
\end{figure*}

\section{Results}

We analyse $A$ and $D$ in the OAM qubit Bloch sphere defined by the bases of Eq.~\eqref{eq:bases}. The singlet of Eq.~\eqref{eq:AD-target} is perfectly anti-correlated in every common local basis, in particular in $Z$ and $X$. From the fourfold coincidences $N(b_j,b_k)$, the number of events in which photons $A$ and $D$ are projected onto the states $\ket{b_j}$ and $\ket{b_k}$ of the same basis $\mathcal{B}$ ($j,k\in\{0,1\}$), we build the normalized two-outcome correlation matrices $P^{(Z)}$ and $P^{(X)}$, with elements $P(b_j,b_k)=N(b_j,b_k)/\sum_{j',k'}N(b_{j'},b_{k'})$, and the anti-diagonal weights
\begin{align}
    C_Z&=P(z_0,z_1)+P(z_1,z_0),\nonumber\\
    C_X&=P(x_0,x_1)+P(x_1,x_0).
    \label{eq:CZCX}
\end{align}
For any separable state measured in two MUBs, $C_Z+C_X\le 1+1/d$, so that with $d=2$ the witness $W_2=C_Z+C_X$ obeys $W_2\le3/2$, while the ideal singlet reaches $W_2=2$ \cite{spengler2012MUBs}; a measured $W_2>3/2$ therefore certifies entanglement in the analysed OAM subspace (Sec.~\ref{sec:methods}).

\begin{figure}[t]
\centering
\includegraphics[width=\columnwidth]{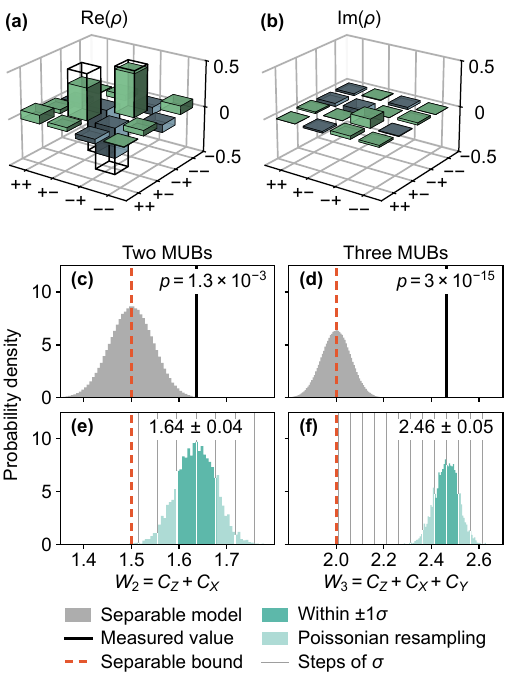}
\caption{\textbf{Entanglement certification of the swapped state.} (a),(b) Real and imaginary parts of the density matrix reconstructed by MLE, in the basis $\{\ket{\pm1}_A\ket{\pm1}_D\}$; the black frames mark the ideal singlet. (c),(d) Witnesses $W_2$ and $W_3$ for $10^6$ simulated realizations of the least favourable separable state; $p$ is the probability that it reaches the measured value. (e),(f) Poissonian resampling of the measured counts; the thin lines are spaced by one standard deviation $\sigma$.}
\label{fig:analisis}
\end{figure}

The data are summarized in Figs.~\ref{fig:setup-data}(b,c) and~\ref{fig:analisis}. Over approximately 100~h of acquisition we recorded 1562 fourfold events across the 36 combinations of the six projectors of Eq.~\eqref{eq:bases} on each photon [Fig.~\ref{fig:setup-data}(b)], which also allow the full reconstruction of the state discussed below [Fig.~\ref{fig:analisis}(a,b)]; of these, 240, 144 and 157 events fall in the $Z$, $X$ and $Y$ blocks, respectively, which are the only ones needed for the witness [Fig.~\ref{fig:setup-data}(c)]. We obtain $C_Z=0.82\pm0.02$ and $C_X=0.82\pm0.03$, and thus
\begin{equation}
    W_2=C_Z+C_X=1.64\pm0.04>3/2,
\end{equation}
where the uncertainties follow from Poissonian counting statistics (Sec.~\ref{sec:methods}); the corresponding distribution of $W_2$ is shown in Fig.~\ref{fig:analisis}(e). The measured $W_2$ lies more than three standard deviations above the separable bound. Because the number of events is finite, a separable state on the bound can still fluctuate above it, and how likely this is depends on how the state distributes its correlation among the bases. We therefore compare with the least favourable separable state, i.e., the one most likely to reach the measured value with the same number of events per basis as the experiment [grey distributions in Fig.~\ref{fig:analisis}(c,d); SM, Sec.~\ref{sup:witness}]. For $W_2$ this probability is $p\approx1.3\times10^{-3}$.
To increase the statistical significance, we also use the third MUB available for qubits, $Y$, and the witness $W_3=C_Z+C_X+C_Y$, with separable bound $W_3\le2$ \cite{spengler2012MUBs}. Including the $Y$ block, $C_Y=0.83\pm0.03$, gives $W_3=2.46\pm0.05$, which exceeds the bound by more than nine times its uncertainty [Fig.~\ref{fig:analisis}(f)]. For the least favourable separable state the probability of reaching this value is $p\approx3\times10^{-15}$, and none of $10^6$ simulated realizations, each with the same number of events per basis as the experiment, reached the measured $W_3$ [grey distribution in Fig.~\ref{fig:analisis}(d)]. Correcting for unequal efficiencies of the projectors changes $W_3$ by only 0.002.
\looseness=-1 For the measured correlations, reaching $p<10^{-3}$ requires about 6.5 times fewer events per basis with three bases than with two, i.e., about four times less acquisition time.

\looseness=-1 For comparison, the full set of 36 projections allows a reconstruction of the two-qubit density matrix by maximum-likelihood estimation (MLE) \cite{james2001measurement}, shown in Fig.~\ref{fig:analisis}(a,b). The reconstructed state has a singlet fidelity of $F=0.73\pm0.02$, a purity of $\Tr\rho^2=0.62\pm0.03$ and a concurrence of $\mathcal{C}=0.54\pm0.05$ \cite{wootters1998entanglement}, confirming the entanglement certified by the witness; the uncertainties are obtained by Monte Carlo resampling of the counts (Sec.~\ref{sec:methods}).

\section{Discussion}

The node combines a spin--orbit interface with a polarization HOM measurement to swap entanglement onto the OAM of two photons that never met, and certifies it with only two measurement bases per photon; a third one increases the statistical significance. Since the intermediary photons interfere and are jointly measured in polarization, the parts of the link most affected by mode scrambling can operate on this more resilient DOF, and the spatial entanglement is recovered only at the analysers. In the network of Fig.~\ref{fig:concept}, both channels, the deployed SMF from Alice and the free-space link from Bob, therefore carry only polarization-encoded photons to Charlie, while the OAM photons never leave Alice's and Bob's labs: the swapped OAM entanglement is established between two photons that were never transmitted, and the spatial modes are never exposed to turbulence or to the fibre.

Because the swapped subspace, $\ell=\pm2q$, is set by the topological charge of the $q$-plates, the node is reconfigurable: replacing the plates selects a different OAM subspace while the polarization-based swap, the fibres and the FBS remain unchanged. Since in Fig.~\ref{fig:concept} the $q$-plates sit in the users' labs, the same relay could connect different pairs of users, each working in the OAM subspace set by their plates, as $q$-plates of different charge give access to different OAM subspaces after transport through a single SMF \cite{liu2020multidimensional}. As the $q$-plates are tuned electrically, this selection need not be mechanical: plates of different charge could be stacked in each arm, with only one switched on at a time.

Swapping genuinely high-dimensional entanglement requires a high-dimensional joint measurement. Linear-optics schemes with ancillary photons have recently been proposed for this task in any photonic DOF \cite{baghdasaryan2026entanglement}. Combining them with a spin--orbit interface would require transcribing a $d$-dimensional OAM state onto a $d$-dimensional fibre-compatible encoding, such as polarization combined with time bins, so that both the channel and the joint measurement remain robust. The certification strategy already scales: witnesses built from two MUBs certify entanglement, and a lower bound on its Schmidt number, in arbitrary dimension \cite{spengler2012MUBs,bavaresco2018measurements,morelli2023resource}. With single-outcome projections, as implemented here, they use $2d^2$ joint settings, whereas FST in all $d+1$ MUBs uses $[d(d+1)]^2$ (8 versus 36 for $d=2$), which quickly becomes impractical at the fourfold rates of a swapping experiment \cite{thew2002qudit}.

\looseness=-1 The fidelity and fourfold rate of the present implementation can be improved on several fronts. The $50\%$ loss intrinsic to the $q$-plate/SMF filtering of each intermediary photon is independent of the OAM value and therefore reduces only the fourfold rate, by a factor of four. This is the ideal value: reflections at the $q$-plates and the coupling of the converted mode into the SMF, which depends on $|\ell|$, add further losses, equal for $\ell=+1$ and $\ell=-1$ when the plates are well centred. Multi-pair emission from a single pumped crystal and residual spectral and polarization distinguishability at the FBS, which limit the fidelity, can be addressed with independent, spectrally engineered sources and active polarization stabilization. For the elementary link itself, two of the requirements that matter for a repeater node -- a carrier for the joint measurement that survives standard infrastructure, and compatibility with deployed SMF -- are already met by design; the next step is testing the node with the intermediary photons routed through deployed telecom fibre, and eventually over a turbulent free-space channel, to quantify the visibility and fidelity of the swapped state under those conditions. In this proof of principle both pairs were generated at 810~nm by degenerate down-conversion, a choice made for experimental convenience. Changing the phase-matching condition of the source to non-degenerate down-conversion would place the intermediary photons $B$ and $C$ in the telecom band for deployed SMF, or in a suitable atmospheric window for free-space links, while $A$ and $D$ remain at wavelengths suited to the SLMs and silicon detectors. As the $q$-plates are tuned electrically, they can be adjusted to the new wavelength, so that only the fibre components and the detectors of the BSM would have to be exchanged for their telecom counterparts.

\section{Methods}
\label{sec:methods}

\subsection{Experimental setup}
\label{sec:setup-methods}
Figure~\ref{fig:setup-data}(a) shows a simplified view of the setup; the detailed layout is given in the SM, Sec.~\ref{sup:setup}. In this proof of principle the stations of Alice, Bob and Charlie share one optical table, one crystal and one SLM, and the coloured regions mark the paths that belong to each of them. A single beta-barium borate crystal pumped at 405~nm produces the two SPDC pairs, which are collected at four points of the emission ring: diametrically opposite points give one pair, $A$--$B$ for Alice and $C$--$D$ for Bob. After a 10~nm bandpass filter and a lens, three D-shaped mirrors separate the four paths. Photons $A$ and $D$ are projected by the two halves of the SLM and coupled into SMFs. Photons $B$ and $C$ pass through the $q$-plates, with a motorized delay line in arm $C$ for temporal overlap, and are coupled into the two input fibres of the FBS; in the figure, $B$ reaches Charlie's region in fibre and $C$ in free space, which stands for the two links of Fig.~\ref{fig:concept}. The four detector signals are recorded by the time tagger. Before the swapping, the temporal, spectral and polarization overlap of $B$ and $C$ at the FBS is optimized through the delay line, spectral filtering to 3~nm on $B$ and $C$, and polarization control in the free-space section of arm $C$, respectively (SM, Sec.~\ref{sup:hom}); the bottom-left inset shows the best HOM interference achieved without $q$-plates, i.e., for $\ell=0$ in all paths, with a fitted visibility of $0.90\pm0.02$. With the $q$-plates on, the $Z$ block of the tomography sets a lower bound of $0.78\pm0.04$ on this visibility (SM, Sec.~\ref{sup:hom}).

\subsection{Initial states and spin--orbit map}
In the OAM qubit sector the source states are
\begin{align}
    \ket{\Psi}_{AB}&=\tfrac{1}{\sqrt2}\!\left(\km_A\kp_B+e^{i\theta_{AB}}\kp_A\km_B\right),\nonumber\\
    \ket{\Psi}_{CD}&=\tfrac{1}{\sqrt2}\!\left(\kp_C\km_D+e^{i\theta_{CD}}\km_C\kp_D\right),
    \label{eq:sources}
\end{align}
where the anti-correlation in $\ell$ follows from OAM conservation in SPDC with a Gaussian pump, and $\theta_{AB},\theta_{CD}$ absorb source, propagation and SLM conventions. A $q=1/2$ plate acts as in Eq.~\eqref{eq:qplate-ladder}. The photons leave the crystal horizontally polarized, i.e., in an equal superposition of $\ket{R}$ and $\ket{L}$; for an input $\kp$ only the $\ket{R}$ component is converted to $\ell=0$ and survives the SMF as $\ket{L}$, while for $\km$ only the $\ket{L}$ component survives, emerging as $\ket{R}$. This is the map of Eq.~\eqref{eq:map}, with a $50\%$ filtering loss per photon that is independent of the OAM value and therefore does not bias the state. Writing $\varphi_{AB}=\theta_{AB}+\chi$ and $\varphi_{CD}=\theta_{CD}+\chi$, with $\chi$ the phase of Eq.~\eqref{eq:map}, and absorbing into them any other stable phase, the four-photon state before the FBS is
\begin{align}
\ket{\Psi}_{ABCD}=\tfrac12\big(&
   \km_A\km_D\,\ket{LL}_{BC}\nonumber\\
 +&\,e^{i\varphi_{CD}}\km_A\kp_D\,\ket{LR}_{BC}\nonumber\\
 +&\,e^{i\varphi_{AB}}\kp_A\km_D\,\ket{RL}_{BC}\nonumber\\
 +&\,e^{i(\varphi_{AB}+\varphi_{CD})}\kp_A\kp_D\,\ket{RR}_{BC}\big).
\label{eq:four}
\end{align}

\subsection{FBS coincidence projection}
With FBS input operators $b_\sigma^\dagger$, $c_\sigma^\dagger$ and output operators $u_\sigma^\dagger$, $v_\sigma^\dagger$ ($\sigma=R,L$), the balanced relations $b_\sigma^\dagger\to(u_\sigma^\dagger+v_\sigma^\dagger)/\sqrt2$ and $c_\sigma^\dagger\to(u_\sigma^\dagger-v_\sigma^\dagger)/\sqrt2$ retain, for one-photon-per-output events, only the antisymmetric input component. A coincidence therefore corresponds, up to the detection efficiency, to the projector $\Pi^-_{BC}=\ket{\Psi^-}\!\bra{\Psi^-}_{BC}$ with $\ket{\Psi^-}_{BC}=(\ket{R}_B\ket{L}_C-\ket{L}_B\ket{R}_C)/\sqrt2$. Only the $\ket{RL}_{BC}$ and $\ket{LR}_{BC}$ terms of Eq.~\eqref{eq:four} have antisymmetric support, and projecting yields
\begin{equation}
    \ket{\psi}_{AD}\propto\kp_A\km_D-e^{i\theta}\km_A\kp_D,
\end{equation}
with $\theta=\varphi_{CD}-\varphi_{AB}+\theta_{\rm FBS}$, where $\theta_{\rm FBS}$ is the residual relative phase between $\ket{R}$ and $\ket{L}$ acquired by photon $C$ with respect to photon $B$ in the input fibres and the FBS; phase compensation, performed once during calibration, gives the target singlet of Eq.~\eqref{eq:AD-target}. The data support this single calibration: a drift of $\theta$ during the acquisition would lower $C_X$ and $C_Y$ but not $C_Z$, whereas the three correlations agree within their uncertainties [Fig.~\ref{fig:setup-data}(c)], as expected for partial distinguishability at the FBS, which degrades all bases alike. Equation~\eqref{eq:four} also shows how the projections of $A$ and $D$ remotely set the FBS inputs: projecting them onto $\kp\kp$ or $\km\km$ selects the $\ket{RR}$ or $\ket{LL}$ terms, which are symmetric and bunch, whereas $\kp\km$ and $\km\kp$ select the terms with antisymmetric support that produce coincidences.

\subsection{Entanglement witness}
\looseness=-1 Normalizing the fourfold counts within each $2\times2$ block of settings, the ideal probabilities for the OAM projections $\ket{\mu_j}$ of photon $A$ and $\ket{\nu_k}$ of photon $D$, with $\mu,\nu\in\{z,x,y\}$ and $j,k\in\{0,1\}$, read $P(\mu_j,\nu_k)=\left(1+\delta_{\mu\nu}\left(1-2\delta_{jk}\right)\right)/4$. Defining the correlation in a basis $\mathcal{B}=\{\ket{b_0},\ket{b_1}\}$ as $C_{\mathcal{B}}=\sum_j P(b_j,\pi_{\mathcal{B}}(b_j))$, where $\pi_{\mathcal{B}}$ is the target outcome pairing (for the singlet, the swap of the two outcomes in each common local basis), any separable state measured in two MUBs of local dimension $d$ obeys $C_Z+C_X\le 1+1/d$ \cite{spengler2012MUBs}. With $d=2$ this yields $W_2=C_Z+C_X\le3/2$, so that $W_2>3/2$ certifies entanglement in the two-dimensional OAM subspace, the ideal singlet saturating $C_Z=C_X=1$, i.e., $W_2=2$. For qubits all three MUBs are accessible, and the sum $W_M$ of the correlations in $M$ bases obeys $W_M\le 1+r(M-1)/d$ for a state of Schmidt number $r$ \cite{spengler2012MUBs,morelli2023resource}, giving $W_3=C_Z+C_X+C_Y\le2$ for separable states ($r=1$, $M=3$) and $W_3=3$ for the ideal singlet. The circular block $C_Y$ thus serves both as a consistency check and as a strengthening of the witness. The definitions of Eq.~\eqref{eq:CZCX} do not require the blocks to be measured with the same integration time, since each is normalized separately. Like any witness evaluated on detected events, it assumes well-characterized projectors and fair sampling.

\subsection{Data acquisition and processing}
Each of the 36 projection settings of Fig.~\ref{fig:setup-data}(b) was integrated for 60~s per cycle, and the full set of settings was cycled repeatedly, with the fourfold counts of each setting summed over the cycles, for a total acquisition time of approximately 100~h. The analysis assumes Poissonian counting statistics, which is supported by the data: for one of the acquisition runs (52 cycles, approximately 32~h), the cycle-to-cycle variance of the fourfold counts of each setting is consistent with their mean, with an average Fano factor of 0.98. Propagating the Poissonian uncertainties of the anti-correlated and correlated counts in each basis $\mathcal{B}$ \cite{bevington2003data} gives $\sigma_{C_{\mathcal{B}}}=\left(C_{\mathcal{B}}(1-C_{\mathcal{B}})/n_{\mathcal{B}}\right)^{1/2}$, with $n_{\mathcal{B}}$ the number of fourfold events in basis $\mathcal{B}$; since each basis is measured on a separate set of events, the uncertainties of $W_2$ and $W_3$ are the quadrature sums of those of their terms. The distributions of $W_2$ and $W_3$ shown in Fig.~\ref{fig:analisis}(e,f) are obtained by Monte Carlo resampling, drawing each of the 36 raw counts from a Poisson distribution with mean equal to the measured value (5000 samples); their standard deviations reproduce the analytical uncertainties to the quoted precision. All histograms of Fig.~\ref{fig:analisis} are normalized to unit area and share a common bin width of 0.0056. The density matrix is reconstructed by MLE with the standard Cholesky parametrization \cite{james2001measurement}, the fidelity is $F=\bra{\Psi^-}\rho\ket{\Psi^-}$, the purity $\Tr\rho^2$, and the concurrence is computed from the eigenvalues of $\rho(\sigma_y\otimes\sigma_y)\rho^*(\sigma_y\otimes\sigma_y)$ \cite{wootters1998entanglement}; since these quantities depend nonlinearly on all 36 counts through the reconstruction, their uncertainties are the standard deviations over 1000 reconstructions from the same resampling. The statistical significance of the violation is further assessed against the separable states that saturate the bound, taking the least favourable one, as described in the SM, Sec.~\ref{sup:witness}.

\subsection{Experimental parameters}
\looseness=-1 The two photon pairs are generated by pumping a 1-mm-thick beta-barium borate crystal with a frequency-doubled Ti:sapphire laser at 405~nm (520~mW, 80~MHz repetition rate, i.e., 12.5~ns period), producing degenerate photons at 810~nm. All four photons are filtered with 10~nm bandpass filters centred at 810~nm, and the intermediary photons $B$ and $C$ are additionally filtered with 3~nm bandpass filters centred at 809~nm to enhance their spectral indistinguishability at the FBS. All arms are coupled into standard SMFs. Detection is performed by single-photon counting modules (nominal peak detection efficiency of 65\%), while time tagging and coincidence counting are handled by a time tagger set with a coincidence window of 1~ns (34~ps RMS jitter). Averaged over a 32-h run of the tomography, the twofold coincidence rates between $A$ and the two FBS outputs were 588 and 546~s$^{-1}$, and those between $D$ and the two outputs 477 and 516~s$^{-1}$. The 1562 fourfold events of the tomography, recorded over approximately 100~h, correspond to a mean fourfold rate of about 16 per hour. The Ti:sapphire pulses last about 140~fs, and events with three or more pairs, estimated from the singles and twofold rates of the tomography, account for about 11\% of the fourfold events. As the two pairs occupy four distinct modes, two-pair events factorize into $\ket{\Psi}_{AB}\otimes\ket{\Psi}_{CD}$ [Eq.~\eqref{eq:sources}], unlike four photons split from the same two modes, whose post-selected state is genuinely multipartite entangled \cite{hiesmayr2016observation}.

\section*{Data availability}
The data supporting this study are available from the corresponding author upon reasonable request.

\begin{acknowledgments}
AV acknowledges financial support from the Ram\'on y Cajal Fellowship RYC2023-043066-I, and Grant Nos. \mbox{PID2024-160393NB-I00} and \mbox{PID2024-156240OB-C22} funded by MICIU/AEI/10.13039/501100011033 and FSE+. LM, BP and FC acknowledge support from the European Research Council (ERC), Advanced Grant No.~694683 (PHOSPhOR). JL acknowledges financial support from the Engineering and Physical Sciences Research Council via grants EP/Z533178/1 and EP/Z533166/1.
\end{acknowledgments}

\section*{Competing interests}
The authors declare no competing interests.

\bibliography{mybibfile}

\onecolumngrid 
\clearpage
\setcounter{section}{0}
\setcounter{equation}{0}
\setcounter{figure}{0}
\setcounter{table}{0}
\renewcommand{\thesection}{S\arabic{section}}
\renewcommand{\theequation}{S\arabic{equation}}
\renewcommand{\thefigure}{S\arabic{figure}}
\renewcommand{\thetable}{S\arabic{table}}
\def\theHsection{S\arabic{section}}
\def\theHequation{S\arabic{equation}}
\def\theHfigure{S\arabic{figure}}
\def\theHtable{S\arabic{table}}

\begin{center}
{\large\bfseries Supplemental Material}\\[5pt]
{\normalsize Entanglement swapping through a spin--orbit hybrid qubit interface}\\[6pt]
{\small Javier~Fern\'andez,$^{1,*}$ Philipp~Oleynik,$^{2,*}$ Kiki~Dekkers,$^{2}$ Maddox~Aherin,$^{2}$ Lorenzo~Marrucci,$^{3}$ Bruno~Piccirillo,$^{3}$ Filippo~Cardano,$^{3}$ Gerard~J.~Machado,$^{4}$ Adam~Vall\'es,$^{1}$ Andrew~Forbes,$^{5}$ and Jonathan~Leach$^{2}$}\\[4pt]
{\footnotesize\itshape
$^{1}$Departament de F\'isica, Universitat Aut\`onoma de Barcelona, E-08193 Bellaterra, Spain\\
$^{2}$IPaQS, SUPA, Heriot-Watt University, Edinburgh EH14 4AS, United Kingdom\\
$^{3}$Dipartimento di Fisica, Universit\`a di Napoli Federico II, Complesso Universitario di Monte Sant'Angelo, Napoli, Italy\\
$^{4}$Clarendon Laboratory, University of Oxford, Parks Road, Oxford OX1 3PU, United Kingdom\\
$^{5}$School of Physics, University of the Witwatersrand, Johannesburg 2050, South Africa\\[2pt]
\upshape$^{*}$These authors contributed equally to this work.}
\end{center}

\section{Experimental setup}
\label{sup:setup}
\begin{figure}[!b]
\centering
\includegraphics[width=\linewidth]{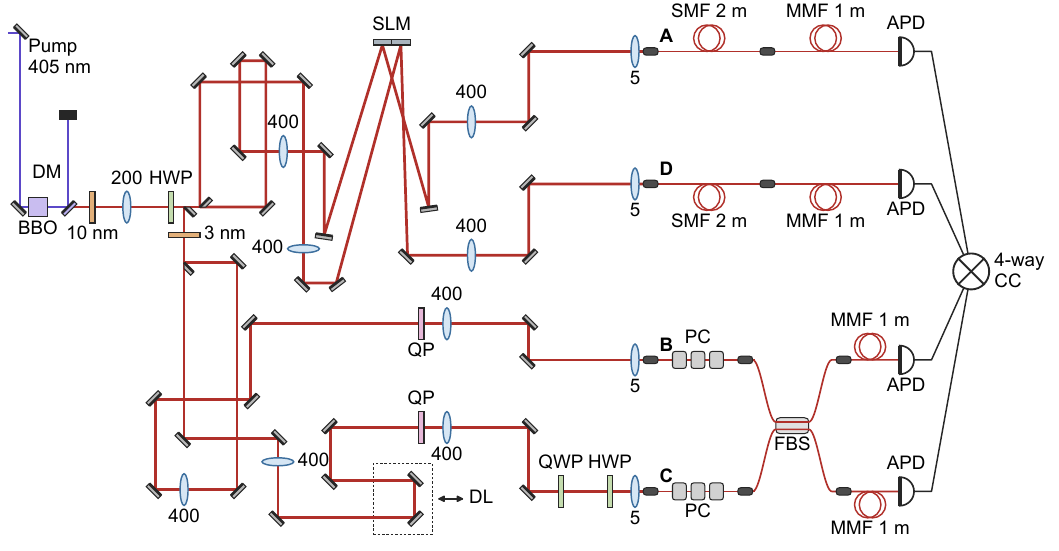}
\caption{\textbf{Detailed layout of the setup} (not to scale). The folds of the four paths follow the laboratory layout, so that the mirror positions can be reproduced; the mirrors with one end cut parallel to the passing beam are the D-shaped mirrors that separate the paths. Numbers next to the lenses are focal lengths in mm. BBO: beta-barium borate crystal. DM: dichroic mirror that removes the pump. 10~nm and 3~nm: bandpass filters centred at 810 and 809~nm, respectively. SLM: spatial light modulator, divided into two halves, with photon $A$ aimed at the centre of one and photon $D$ at the centre of the other. QP: $q$-plate. DL: delay line on a translation stage. HWP, QWP: half- and quarter-wave plates (a HWP on the common beam before the first D-shaped mirror, and a QWP and a HWP for the polarization control of photon $C$). SMF, MMF: single-mode and multimode fibres; the dark rounded symbols are fibre couplers and fibre-to-fibre connectors. PC: fibre polarization controller. FBS: fibre beam splitter. APD: avalanche photodiode detector. 4-way CC: fourfold coincidence counting in the time tagger.}
\label{fig:setup-detailed}
\end{figure}

\looseness=-1 Figure~\ref{fig:setup-detailed} shows the layout of the final configuration, with the focal lengths of all lenses, the positions of the spectral filters and the fibre connections. A beta-barium borate crystal pumped at 405~nm produces the two SPDC pairs, which are collected along the four paths $A$--$D$ after a dichroic mirror that removes the pump, a 10~nm bandpass filter, a 200~mm lens and a half-wave plate. Photons $A$ and $D$ are projected by the SLM and coupled into SMFs. Photons $B$ and $C$ pass through the $q$-plates, with a delay line in arm $C$ for temporal overlap and a quarter- and half-wave plate pair for polarization control, and are coupled into the two input fibres of the FBS; polarization paddles on the fibre act as additional polarization control. The four detector signals are recorded by the time tagger. Figure~\ref{fig:setup-data}(a) of the main text is a simplified view of this layout.

\section{Alignment and calibration procedure}
\label{sup:align}

\subsection{Selection of the two photon pairs}
\looseness=-1 The two photon pairs are extracted from the SPDC emission of a single crystal. Using software developed in-house, we acquire the image of the down-conversion ring on a camera and automatically select the four positions on its two diagonals at which the brightness is maximized [top-left inset of Fig.~\ref{fig:setup-data}(a) of the main text]. These four positions define the paths $A$, $B$, $C$ and $D$, so that two mutually uncorrelated pairs, $A$--$B$ and $C$--$D$, are obtained from the same crystal.

\subsection{Calibration of the SLM projectors}
\begin{figure}[!ht]
\centering
\includegraphics[width=0.8\linewidth]{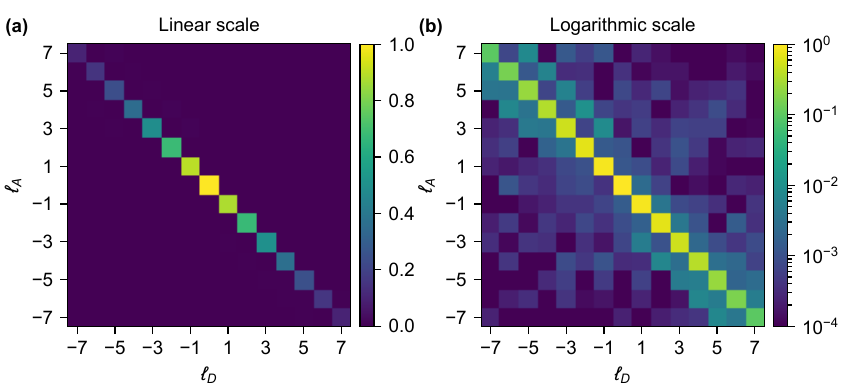}
\caption{\looseness=-1 \textbf{Classical calibration of the SLM projectors.} Power coupled into the fibre of arm $D$ when 810~nm light is injected through arm $A$ with a mirror at the crystal plane, as a function of the OAM values $\ell_A$ and $\ell_D$ displayed on the SLM, normalized to the $\ell_A=\ell_D=0$ element, in (a) linear and (b) logarithmic scale. The mean nearest-neighbour crosstalk around $\ell=0$ is 0.14\%.}
\label{fig:slm}
\end{figure}

To characterize the OAM projectors on $A$ and $D$ independently of the source, we replace the crystal by a mirror and send classical light at 810~nm backwards through arm $A$, measuring the power coupled into the fibre of arm $D$ for every pair of OAM holograms $\ell_A,\ell_D\in\{-7,\dots,7\}$ displayed on the SLM. The reflection at the mirror plane reproduces the OAM anti-correlation of the source, so that the transmitted light appears on the anti-diagonal $\ell_D=-\ell_A$ (Fig.~\ref{fig:slm}). With 0.443~mW of injected light, the coupling efficiency between the two arms reaches 90.4\%. We quantify the crosstalk as the mean of the four nearest-neighbour elements of the $\ell=0$ element relative to the $\ell=0$ element itself, obtaining 0.14\%. For the $\ell=\pm1$ subspace used in the experiment, the two transmitted anti-diagonal elements differ by only 2\%, so that the projectors do not introduce a significant bias between the two basis states. This classical measurement shows the level of performance the SLM projectors can reach, independently of the photon statistics of the four-photon experiment.

\subsection{$q$-plate characterization}
\begin{figure}[!ht]
\centering
\begin{minipage}[c]{0.56\linewidth}
\includegraphics[width=\linewidth]{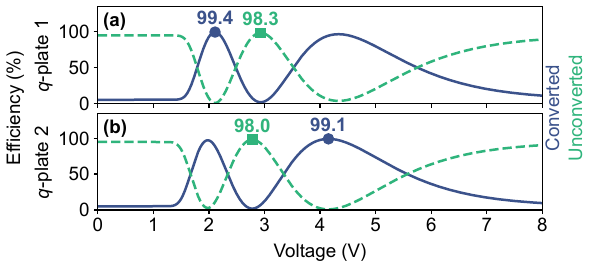}
\end{minipage}\hfill
\begin{minipage}[c]{0.42\linewidth}
\centering\small
\begin{tabular}{ccccc}
\hline
Voltage & State & Efficiency & $P_H$ & $P_V$ \\
(V) & & (\%) & ($\mu$W) & ($\mu$W) \\
\hline
\multicolumn{5}{l}{$q$-plate 1} \\
\textbf{2.11} & C & 99.4 & 1.2 & 205 \\
2.93 & U & 98.3 & 205 & 3.5 \\
\hline
\multicolumn{5}{l}{$q$-plate 2} \\
2.78 & U & 98.0 & 256 & 5.1 \\
\textbf{4.15} & C & 99.1 & 2.3 & 263 \\
\hline
\end{tabular}
\end{minipage}
\caption{\textbf{Tuning of the $q$-plates.} (a),(b) Efficiency of conversion (solid) and fraction of unconverted light (dashed) as a function of the applied voltage for $q$-plates 1 and 2, i.e., the fraction of the power in the V and H outputs of the PBS, respectively. The markers are the measured settings; the curves are a schematic of the behaviour observed while searching for these maxima and minima, drawn through them. Table: powers $P_H$ and $P_V$ in the two PBS outputs and efficiency at the measured settings (C, converted; U, unconverted), with the voltages used in the experiment in bold.}
\label{fig:qp}
\end{figure}

Equation~\eqref{eq:qplate-ladder} describes the action of a $q$-plate of charge $q$; when tuned, it converts the spin fully, while a partially tuned or untuned $q$-plate only converts some light.
The $q$-plates were tuned by applying voltage, such that the phase retardation was as close as possible to an odd multiple of $\pi$. Voltage was applied as an AC square wave. Polarization was used to characterize the conversion efficiency using laser illumination at the SPDC wavelength. Horizontally polarized light (H) was incident on the plate, placed between two quarter-wave plates (QWPs), followed by a polarizing beam splitter (PBS). QWPs were set at $45^\circ$ to the optic axis and power was measured after the PBS. Figure~\ref{fig:qp} summarizes the characterization. Above a threshold voltage, the retardance of the liquid-crystal plates decreases monotonically with the applied voltage, so that the conversion efficiency oscillates: it has two maxima, at retardances $3\pi$ and $\pi$, separated by the unconverted setting at $2\pi$, which lies at similar voltages for the two plates (2.93 and 2.78~V). The two maxima of each plate differ slightly in efficiency, and we chose the higher one: the $3\pi$ maximum for $q$-plate~1 (2.11~V, 99.4\%) and the $\pi$ maximum for $q$-plate~2 (4.15~V, 99.1\%). The different optimal voltages therefore correspond to different retardance orders, equivalent for the conversion. For the swapping experiment we always prepare H polarized photons (as generated by SPDC), which were then accepted or rejected by the single-mode fibre depending on their initial OAM index.

\looseness=-1 For an ideal $q$-plate, polarization is perfectly coupled to OAM mode conversion; however, the OAM conversion efficiency may not be the same as the polarization one; nevertheless, the above gives the optimal voltage settings required.

\looseness=-1 An imbalance between the $\ell=+1$ and $\ell=-1$ channels (SLM projectors, $q$-plates or fibre coupling) that changes the weights of the two terms of the swapped state would unbalance the two anti-correlated entries of the $Z$ block [Fig.~\ref{fig:setup-data}(b) of the main text]. These are 111 and 85 events, a difference of 1.9 standard deviations that, even if entirely due to such an imbalance, would lower the singlet fidelity by only 0.4\%. The different retardance orders of the two plates, $3\pi$ and $\pi$, do not make the converted photons appreciably distinguishable: with the retardance inversely proportional to the wavelength, the conversion efficiency stays within $10^{-4}$ of its maximum across the 3~nm band, and the $2\pi$ difference in retardance shifts the optical path of the converted light by half a wavelength, 1.4~fs, negligible against the 0.6~ps width (FWHM) of the HOM dip.

\section{Temporal and polarization overlap at the FBS}
\label{sup:hom}
\looseness=-1 Before the swapping, the HOM interference of $B$ and $C$ at the FBS must reach maximum visibility. Temporal, spectral and polarization overlap are ensured through a delay stage in path $C$, additional spectral filtering to 3~nm on $B$ and $C$, and polarization control, respectively; the delay-stage scan is performed first. The bottom-left inset of Fig.~\ref{fig:setup-data}(a) of the main text shows the best interference achieved for $\ell=0$ photons: the fourfold coincidences as a function of the time delay $\tau$ set by the delay stage, with the $q$-plates on $B$ and $C$ switched off and the SLMs on $A$ and $D$ displaying a blazed grating ($\ell=0$ projection). Each point was integrated for 40~s over 20 repetitions of the full scan range, i.e., approximately 13~minutes per point, and the flipped-Gaussian fit gives a visibility of $0.90\pm0.02$.

\begin{figure}[!ht]
   \centering
   \includegraphics[width=0.8\linewidth]{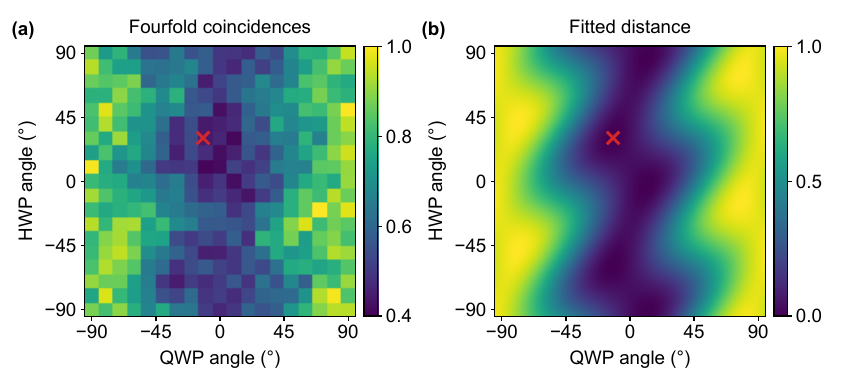}
   \caption{\looseness=-1 Polarization optimization at the FBS. (a) Fourfold coincidences, normalized to their maximum, as a function of the quarter- and half-wave plate angles (QWP and HWP) in arm $C$, relative to the centre of the scan (180~s per point, only the 10~nm filters on all photons). (b) Normalized Poincar\'e-sphere distance $(1-\mathbf{s}_B\cdot\mathbf{s}_C)/2$ between the Stokes vectors of photons $B$ and $C$, obtained from a fit of the data in (a). The red cross marks the minimum of the fitted map, at the same angles in both panels.}
   \label{fig:HOMpol}
\end{figure}

The low fourfold coincidence rate made manual alignment of the fibre paddles impractical, so they were fixed in place, and the polarization overlap was optimized instead with two wave plates of retardances $\lambda/2$ and $\lambda/4$ in the free-space path of photon $C$ (the choice between $B$ and $C$ is arbitrary). With the temporal and spectral overlap already set by the delay-stage scan and the bandpass filters, the visibility is proportional to the polarization overlap $(1+\mathbf{s}_B\cdot\mathbf{s}_C)/2$, where $\mathbf{s}_B$ and $\mathbf{s}_C$ are the normalized Stokes vectors of the two photons. Figure~\ref{fig:HOMpol}(a) shows the normalized fourfold coincidences as a function of the two wave plate angles (in this scan all four photons were filtered only with the 10~nm filters, without the additional 3~nm filters on $B$ and $C$).

The SPDC photons leave the crystal horizontally polarized, so that to a small error both $B$ and $C$ enter their fibres in $\ket{H}$. The fibre arms before the FBS then impart unknown unitaries to the polarizations, of which only the relative one matters for the indistinguishability, and the wave plates compensate it. Scanning both wave plates and fitting the fourfold coincidences gives Fig.~\ref{fig:HOMpol}(b), a map of the Poincar\'e-sphere distance between the polarization vectors of $B$ and $C$. The visibility of this scan is lower than that of the final dip because of the weaker spectral filtering. The wave plates were then fixed, for the rest of the experiment, at the settings that minimize the fourfold coincidences, marked by the red cross in Fig.~\ref{fig:HOMpol}. The HOM alignment was not re-checked during the $\approx100$~h tomography: it was the most stable part of the setup, more so than the fibre coupling and the frequency doubling of the pump laser, and the cycle-to-cycle fourfold counts show no fluctuations beyond Poissonian ones (Fano factor 0.98; Sec.~\ref{sec:methods}).

With the $q$-plates on, the visibility can be read from the $Z$ block of the tomography [Fig.~\ref{fig:setup-data}(b) of the main text]. Equal OAM values on $A$ and $D$ select $\ket{RR}_{BC}$ or $\ket{LL}_{BC}$ [Eq.~\eqref{eq:four} of the main text], which give a coincidence at the FBS with probability $(1-\mathcal{V})/2$, $\mathcal{V}$ being the two-photon interference visibility, whereas opposite values select $\ket{RL}_{BC}$ or $\ket{LR}_{BC}$, which give one with probability $1/2$ for any $\mathcal{V}$. As the four terms have equal weights, the 44 events with equal values and the 196 with opposite ones give $\mathcal{V}=1-44/196=0.78\pm0.04$. This is a lower bound, since multi-pair emission, imperfect conversion and projector crosstalk also add events with equal values.

\section{Full state tomography}
\label{sup:tomo}
On each of $A,D$ we use the six OAM qubit states of the three MUBs $Z$, $X$ and $Y$ of Eq.~\eqref{eq:bases} of the main text. For projectors $\ket{\mu_j}=\alpha_+\kp+\alpha_-\km$ on $A$ and $\ket{\nu_k}=\gamma_+\kp+\gamma_-\km$ on $D$ ($\mu,\nu\in\{z,x,y\}$; $j,k\in\{0,1\}$), the singlet amplitude is
\begin{equation}
    \bra{\mu_j}_A\!\tp\bra{\nu_k}_D\,\ket{\Psi^-}_{AD}
    =\tfrac{1}{\sqrt2}\!\left(\alpha_+^*\gamma_-^*-\alpha_-^*\gamma_+^*\right),
\end{equation}
and $P(\mu_j,\nu_k)$ is its squared modulus. Normalizing by the ideal maximum $1/2$ gives
\begin{equation}
    2P(\mu_j,\nu_k)=\delta_{\mu\nu}\left(1-\delta_{jk}\right)+\tfrac12\left(1-\delta_{\mu\nu}\right),
    \label{eq:singlet-pattern}
\end{equation}
i.e., 1 for opposite outcomes in the same basis, 0 for equal outcomes and $1/2$ across bases: the singlet is anti-correlated within each of the three MUBs and uniform across bases.

To verify the swap we performed FST of the two-qubit state [Fig.~\ref{fig:setup-data}(b) of the main text], which took approximately 100~h, and reconstructed its density matrix by MLE \cite{james2001measurement}. The goal is to certify entanglement between $A$ and $D$, i.e., that an entangled state has been swapped onto them through the BSM on the $B,C$ photons, $\bra{\Psi^-}_{BC}\,\ket{\Psi}_{AB}\otimes\ket{\Psi}_{CD}\propto\ket{\Psi^-}_{AD}$. The reconstructed density matrix [Fig.~\ref{fig:analisis}(a,b) of the main text] confirms it, as its concurrence is greater than zero: it has a purity of $0.62\pm0.03$, a fidelity to $\ket{\Psi^-}\!\bra{\Psi^-}$ of $F=0.73\pm0.02$ and a concurrence of $\mathcal{C}=0.54\pm0.05$, with uncertainties obtained from 1000 Monte Carlo simulations of the MLE for the number of photon detection events, assuming Poissonian counting statistics.

\looseness=-1 FST is time consuming: at low fourfold coincidence rates it is inefficient for demonstrating entanglement even for qubits, while for qudits the polynomial ($\propto d^4$) scaling of the required local filter settings makes it impractical or near impossible \cite{thew2002qudit,bavaresco2018measurements}.

\section{Entanglement witness with two and three MUBs}
\label{sup:witness}
\looseness=-1 The alternative to FST is an entanglement witness: measurements in just two MUBs suffice to certify (high-dimensional) entanglement \cite{bavaresco2018measurements}. With the normalized two-outcome matrices $P^{(Z)}$ and $P^{(X)}$, the anti-diagonal weight is $C_{\mathcal{B}}=\Tr[S\,P^{(\mathcal{B})}]$, $S=\big(\begin{smallmatrix}0&1\\1&0\end{smallmatrix}\big)$, and $W_2=C_Z+C_X$; separable states obey $C_Z+C_X\le1+1/d$, i.e., $W_2\le3/2$ for $d=2$.

For qubits a third MUB is available, and measuring in all three can be advantageous: the bound $W_2\le1+r/d$, with $r$ the Schmidt number ($r=1$ for separable states) and $d$ the dimension of the state \cite{spengler2012MUBs,morelli2023resource}, generalizes to $W_M\le1+r(M-1)/d$, where $M$ is the number of MUBs and $W_M$ the sum of the correlations in them. With limited photon statistics this is preferable to two MUBs, as it lowers the ratio between the correlations achievable by a separable state and by a maximally entangled one.

Let Alice and Bob project photons $A$ and $D$ onto the states $\ket{\mu_j}$ and $\ket{\nu_k}$ of bases $\mu,\nu\in\{z,x,y\}$, with outcomes $j,k\in\{0,1\}$ and joint probability $P(\mu_j,\nu_k)$. Labelling the outcomes so that the target state is correlated in the $M$ common bases, the witness reads
\begin{equation}
W_M=\sum_{\mu=\nu}\;\sum_{j=k}P(\mu_j,\nu_k).
\end{equation}
A maximally entangled state reaches $W_3=3$ through perfect correlations in the three bases $\mu=\nu$, whereas no separable state exceeds $W_3=2$; a simple example on the bound is a classically correlated (separable) model with perfect correlations in the computational basis ($\mu=\nu=z$) and uncorrelated outcomes in the other two, for which $W_3=1+\tfrac12+\tfrac12=2$. From our tomography data, after swapping the labels of the outcomes (the $\ket{\Psi^-}$ state is anti-correlated in all MUBs), we obtain $W_3=2.46\pm0.05$.

With a finite number of events a separable state on the bound can fluctuate above it, so the significance must be assessed against the separable state most likely to reach the measured value; the $Z$-correlated model above lies on the bound but is not that state. A product state with opposite local Bloch vectors $\mathbf{a}$ and $-\mathbf{a}$, $|\mathbf{a}|=1$, gives $C_{\mathcal{B}}=(1+a_{\mathcal{B}}^2)/2$ in each basis, and thus $W_3=2$ for any orientation (and $W_2=3/2$ for $a_Y=0$), with the number of anti-correlated outcomes in each block binomially distributed. These states share the mean value of the witness but not its spread: the statistical fluctuation of each $C_{\mathcal{B}}$, $\left(C_{\mathcal{B}}(1-C_{\mathcal{B}})/n_{\mathcal{B}}\right)^{1/2}$, is largest for weak correlations ($C_{\mathcal{B}}$ close to $1/2$) and for bases with few events. With equal numbers of events per basis the least favourable state therefore shares the correlation equally among the bases, $C_{\mathcal{B}}=3/4$ (two MUBs) or $C_{\mathcal{B}}=2/3$ (three MUBs); with the measured numbers it gives the strongest correlation to the basis with the most events ($Z$) and the weakest to the one with the fewest ($X$). Maximizing over the orientation the probability of reaching the measured value gives $C_Z\approx0.76$, $C_X\approx0.61$ and $C_Y\approx0.63$ for $W_3$, and $C_Z\approx0.83$ and $C_X\approx0.67$ for $W_2$, and the $p$-values follow exactly from the binomial distribution: $p\approx3\times10^{-15}$ for $W_3$ and $p\approx1.3\times10^{-3}$ for $W_2$. A Monte Carlo simulation confirms both: in $10^6$ realizations of each state, each with the same number of events per basis as the experiment, the fraction that reaches the measured $W_2$ agrees with the exact $p$ within its statistical error, and none reaches the measured $W_3$; these realizations are the grey distributions of Fig.~\ref{fig:analisis}(c,d) of the main text. In terms of the Poissonian uncertainty of the measured counts, $W_3$ lies more than nine standard deviations above the separable bound and $W_2$ slightly more than three [Fig.~\ref{fig:analisis}(e,f) of the main text].

The three blocks contain 240, 144 and 157 events for equal integration times, which points to different efficiencies of the projectors. A factor common to a basis cancels in the normalization of its block, but unequal efficiencies within a basis do not. We estimate one factor per projector of $A$ and $D$ from the 24 cross-basis counts, to which the singlet gives equal probabilities [Eq.~\eqref{eq:singlet-pattern}]. These factors are, on average, higher for the pure OAM states of $Z$ than for the superpositions of $X$ and $Y$, predict 240, 149 and 163 events for the three blocks and, divided out, change $W_3$ from 2.464 to 2.466, 28 times less than its uncertainty. As the efficiency of a projector multiplies one anti-correlated and one correlated entry of its block, its effect on $C_{\mathcal{B}}$ largely cancels: an error of 10\% in one projector changes $W_3$ by at most 0.003, and errors of up to 30\% in all 12, in the least favourable combination, by at most 0.13, leaving $W_3\ge2.34$.

\begin{figure}[!hb]
   \centering
   \includegraphics[width=0.8\linewidth]{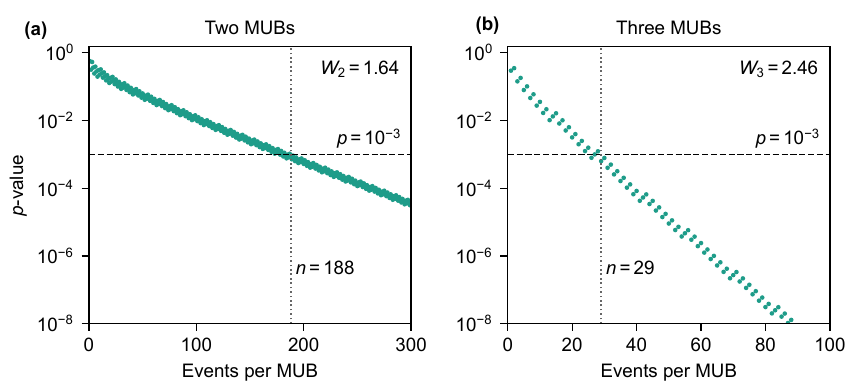}
   \caption{$p$-value against the least favourable separable state, which for equal numbers of events in every MUB has $C_{\mathcal{B}}=3/4$ (two MUBs) and $C_{\mathcal{B}}=2/3$ (three MUBs) in each basis, as a function of the number of events per MUB, computed exactly from the binomial distribution for the measured (a) $W_2=1.64\pm0.04$ and (b) $W_3=2.46\pm0.05$. The dashed lines mark $p=10^{-3}$ and the dotted lines the number of events beyond which $p<10^{-3}$: 188 per MUB with two MUBs and 29 with three.}
   \label{fig:pvalues}
\end{figure}

\looseness=-1 This also sets the required measurement time. Of the 36 settings of the $\approx100$~h tomography, 12 belong to the three MUBs used for the witness. Figure~\ref{fig:pvalues} shows the $p$-value against the least favourable separable state as a function of the number of events per MUB, for the measured correlation strengths: although each measurement cycle is $3/2$ times longer with three MUBs than with two, about six and a half times fewer events per MUB are needed, so that satisfactory statistics are obtained approximately four times faster overall (assuming balanced event rates in each MUB).

\end{document}